\documentclass{IEEEtran}
\usepackage{cite}
\usepackage{amsmath,amssymb,amsfonts}
\usepackage{graphicx}
\usepackage{textcomp,nicefrac}
\usepackage{multirow}
\def\BibTeX{{\rm B\kern-.05em{\sc i\kern-.025em b}\kern-.08em
T\kern-.1667em\lower.7ex\hbox{E}\kern-.125emX}}
\usepackage{scalerel}
\usepackage{tikz}
\usetikzlibrary{svg.path}

\definecolor{orcidlogocol}{HTML}{A6CE39}
\tikzset{
  orcidlogo/.pic={
    \fill[orcidlogocol] svg{M256,128c0,70.7-57.3,128-128,128C57.3,256,0,198.7,0,128C0,57.3,57.3,0,128,0C198.7,0,256,57.3,256,128z};
    \fill[white] svg{M86.3,186.2H70.9V79.1h15.4v48.4V186.2z}
                 svg{M108.9,79.1h41.6c39.6,0,57,28.3,57,53.6c0,27.5-21.5,53.6-56.8,53.6h-41.8V79.1z M124.3,172.4h24.5c34.9,0,42.9-26.5,42.9-39.7c0-21.5-13.7-39.7-43.7-39.7h-23.7V172.4z}
                 svg{M88.7,56.8c0,5.5-4.5,10.1-10.1,10.1c-5.6,0-10.1-4.6-10.1-10.1c0-5.6,4.5-10.1,10.1-10.1C84.2,46.7,88.7,51.3,88.7,56.8z};
  }
}
\newcommand\orcidicon[1]{\href{https://orcid.org/#1}{\mbox{\scalerel*{
\begin{tikzpicture}[yscale=-1,transform shape]
\pic{orcidlogo};
\end{tikzpicture}
}{|}}}}

\usepackage{hyperref}
\hypersetup{
  colorlinks=false,
  linkbordercolor=white,
 urlbordercolor=white,
pdfborder={0 0 0}
}

\begin{document}
\title{Microlens-enhanced SiPMs}

\author{

    G.~Haefeli$^{\textsuperscript{\orcidicon{0000-0002-9257-839X}}}$, F.~Blanc$^{\textsuperscript{\orcidicon{0000-0001-5775-3132}}}$, E.~Currás-Rivera$^{\textsuperscript{\orcidicon{0000-0002-6555-0340}}}$, R.~Marchevski$^{\textsuperscript{\orcidicon{0000-0003-3410-0918}}}$, F.~Ronchetti$^{\textsuperscript{\orcidicon{0000-0003-3438-9774}}}$, O.~Schneider$^{\textsuperscript{\orcidicon{0000-0002-6014-7552}}}$, L.~Shchutska$^{\textsuperscript{\orcidicon{0000-0003-0700-5448}}}$, C.~Trippl$^{\textsuperscript{\orcidicon{0000-0003-3664-1240}}}$, E.~Zaffaroni$^{\textsuperscript{\orcidicon{0000-0003-1714-9218}}}$ and G.~Zunica$^{\textsuperscript{\orcidicon{0000-0002-5972-6290}}}$
    
    \thanks{``This work was supported by the Swiss National Science Foundation under Grant
    215635 and 216604.''}
    \thanks{G.~Haefeli, F.~Blanc, E.~Currás-Rivera, R.~Marchevski, F.~Ronchetti, O.~Schneider, L.~Shchutska, and G.~Zunica are with the Institute of Physics, École Polytechnique Fédérale de Lausanne, Route de Sorge, Lausanne, Switzerland.}
    \thanks{At the time of this work, C.~Trippl was with the Institute of Physics, École Polytechnique Fédérale de Lausanne, and is now at DS4DS, La Salle, Universitat Ramon Llull, Barcelona, Spain.}
    \thanks{At the time of this work, E.~Zaffaroni was with the Institute of Physics, École Polytechnique Fédérale de Lausanne, and is now at Université de Genève, Geneva, Switzerland.}
    
}

\maketitle

\begin{abstract}
A novel concept to enhance the photo-detection efficiency (PDE) of silicon photomultipliers (SiPMs) has been applied and remarkable positive results can be reported. This concept uses arrays of microlenses to cover every second SiPM pixel in a chequerboard arrangement and aims to deflect the light from the dead region of the pixelised structure towards the active region in the centre of the pixel. The PDE is improved up to 24\%, external cross-talk is reduced by 40\% compared to a flat epoxy layer, and single photon time resolution is improved. This detector development is conducted in the context of the next generation LHCb scintillating fibre tracker located in a high radiation environment with a total of 700'000 detector channels. The simulation and measurement results are in good agreement and will be discussed in this work. 
\end{abstract}

\begin{IEEEkeywords}
Cross-talk, Microlens, MLA, PDE, SciFi, SiPM, SPTR
\end{IEEEkeywords}

\section{Introduction}
\label{introduction}
In the context of the scintillating fibre tracker (SciFi Tracker)~\cite{LHCb_U1}, our group, the High Energy Physics Laboratory (LPHE) at EPFL, is working to improve the silicon photomultiplier (SiPM) photodetectors in order to overcome the challenges imposed by radiation, detector size, and the low material budget requirement for the next upgrade in 2034. The ultimate goal of this development is to improve the signal-to-noise ratio. We found that the enhancement of the SiPM with microlenses not only significantly increases the signal but also reduces external cross-talk and has the potential to reduce noise.\par
The choice of the ideal size of the SiPM pixel for a custom design is typically guided by the compromise between the high geometrical fill factor (GFF) and the resulting high photo-detection efficiency (PDE) for large pixels and the manifold disadvantages resulting from higher gain and therefore higher correlated noise, longer recovery time, lower dynamic range, and higher bias current. Adding the radiation environment as an additional criterion, small pixel size and low excess bias voltage (over-voltage, $\Delta V$) are also preferred to reduce the effect of increasing the dark count rate (DCR) (heat dissipation and self-heating), as well as reducing correlated noise. In summary, the small pixel size and the low $\Delta V$ are advantageous in all aspects except for the PDE.\par
The hit detection efficiency of the scintillating fibre tracker technology depends on the actual signal (scintillating light) produced and the PDE of the photodetector. For a given fibre technology and fibre length, stacking several layers of fibres is used to increase the signal for a traversing particle. This comes at the cost of a higher material budget, a large signal spread for incident particles with a large incident angle, and last but not least a higher cost for more scintillating fibres. The implementation of microlens arrays (MLAs) aligned with the pixel structure has been successfully tested, and the results are presented in the following. 

\section{The microlens on SiPMs}
Enhancing imaging photodetectors with pixel level microlenses is commonly used for CCD and CMOS camera sensors~\cite{MLA_general} and has also been successfully applied to SPAD array sensors in the last years~\cite{MLA_SPAD1,MLA_SPAD2,Charbon}. A good overview of the fabrication process of refractive MLAs is given in~\cite{MLA_fabrication1}, focusing on spherical lenses, although cylindrical lenses can be obtained with the same process~\cite{MLA_cylinder}. The process used for our application is based on photoresist processing, followed by a thermal reflow yielding a master mold. This mold is used with a UV-curable polymer to replicate the MLA on the wafer level. Other fabrication processes are microdroplet inkjet printing or MEMS-based. The implementation of microlenses requires high-precision manufacturing, alignment, and a clean-room environment. These types of processes are typically performed by highly specialised research institutions. For this project, the Swiss Center for Electronics and Microtechnology (CSEM)~\cite{CSEM} was sub-contracted for the microlens implementation.\par
\begin{figure}
	\centering 
	\includegraphics[width=0.35\textwidth, angle=0, trim={0 0.5cm 0 0}, clip]{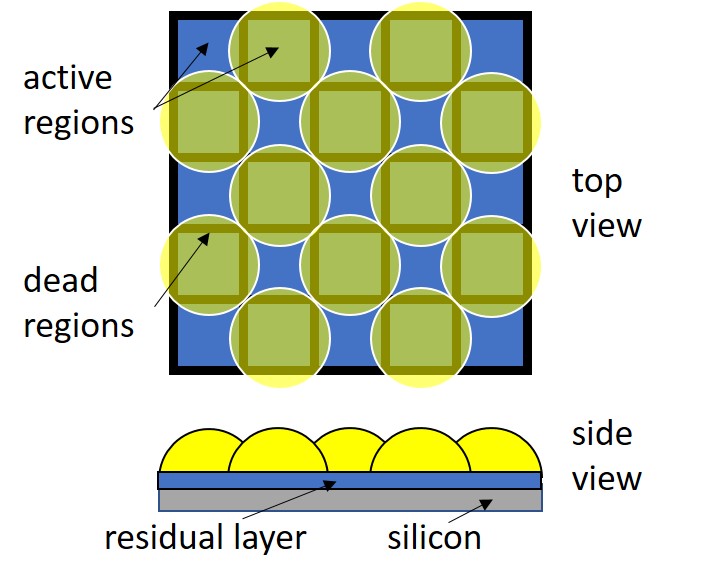}	
	\caption{Schematic representation of the microlens layout. The MLA covers only every second pixel. The advantage of this layout is that the dead region around the pixel can efficiently be avoided by the lens. The remaining region between lenses is covered with a flat residual layer and lies where a fully active region is. } 
	\label{fig:lense_layout}%
\end{figure}

Unlike the typical implementation where the base surface of the lens is square or hexagonal and placed to cover the entire sensor surface, we found an innovative solution by covering every second pixel in a chequerboard structure, as shown in Fig.~\ref{fig:lense_layout}. As the SiPM is operated as a photon counter and not as a pixelised imaging sensor, the deviation of photons from one pixel to a neighbouring pixel is acceptable.

\begin{figure}
\centering
	\includegraphics[width=0.4 \textwidth, angle=0]{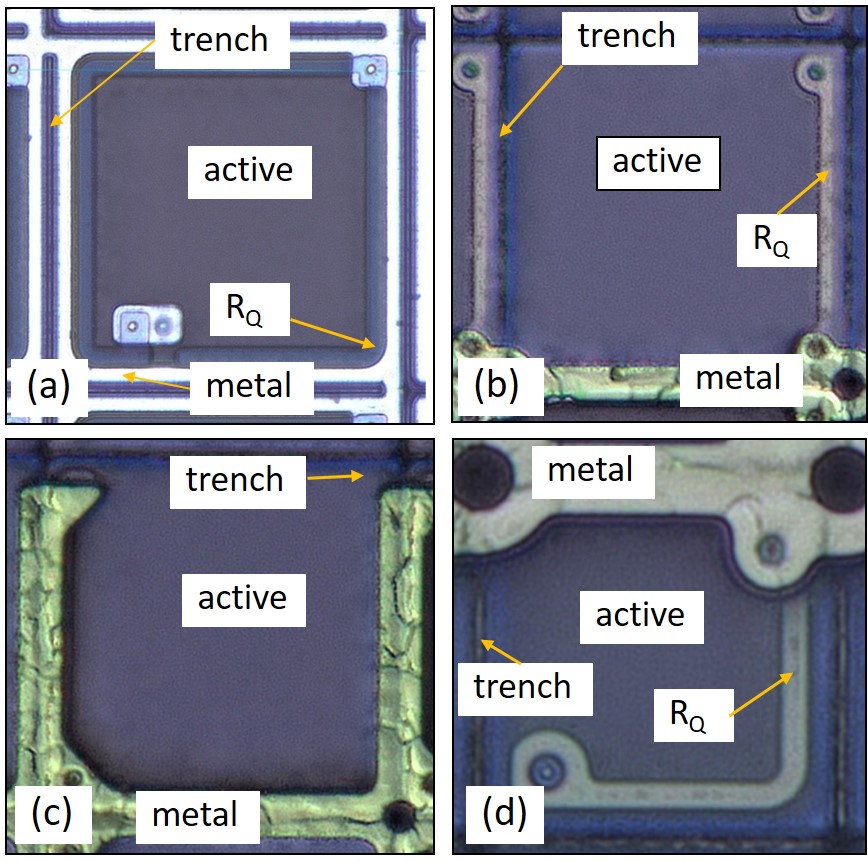}	
	\caption{Microscope pictures of different pixel layouts (a) $42\mu m$ pixel from HPK with thin film metal resistor (70\% transparent), GFF=68.5\% (b) $42\mu m$ pixel from FBK with poly-silicon quench resistor, GFF=81.5\% (c) $31\mu m$ pixel from FBK, GFF=77.7\% (d) $16\mu m$ pixel from FBK, GFF=43.4\%. The ratio between active and total area is compromised by different elements such as trenches, quench resistor, and metal traces.} 
	\label{fig:pixel_layout}
\end{figure}

In this arrangement, the microlens deflects the light by refraction, from the dead border region towards the centre of the lens covered pixel. The area between lenses is covered with a flat region aligned with the active region of adjacent pixels. As illustrated by the pixel layouts in Fig.~\ref{fig:pixel_layout}, the active region is reduced by optical insulation trenches (to suppress optical cross-talk), quench resistors, and metal traces. With an optical microscope, the GFFs of different pixel layouts can be measured and used in the simulation. For the shown examples, the values correspond well with the specifications given by the manufacturers Hamamatsu (HPK)~\cite{Hamamatsu} and Fondazione Bruno Kessler (FBK)~\cite{FBK}. 

With the effect of the microlens, an effective GFF (EGFF) larger than the initial GFF is achieved.\\
The second important modification due to the microlens is related to the concentration of light towards the centre of the pixels. At low $\Delta V$ and therefore low electric field, the region close to the pixel border has a low avalanche trigger probability ($P_{AT}$) and therefore a low PDE. With the microlens, the border region of the pixel is less used (light deviated towards the center) and, as a consequence, the PDE is expected to be enhanced at low $\Delta V$. This is supported by the simulation performed in Ref.~\cite{FBK_LFR} and as well shown in~\cite{LF_efficiency_FBK} by measuring the GFF as a function of $\Delta V$. Furthermore, the time resolution is also improved. In~\cite{Nemallapudi_2016} it is shown that the single photon time resolution (SPTR) is improved for the center region of the pixel. Masking pixel boundaries has been successfully performed by FBK to improve SPTR~\cite{FBK_mask}. With the effect of the lens, improved SPTR for the microlens-enhanced detectors is observed as for the masked structures.\\
The effectiveness of the microlens depends on several factors that are discussed in more detail in the simulation section below. 

\section{Simulation}
\begin{figure}
\centering
    \includegraphics[width=0.235 \textwidth, angle=0]{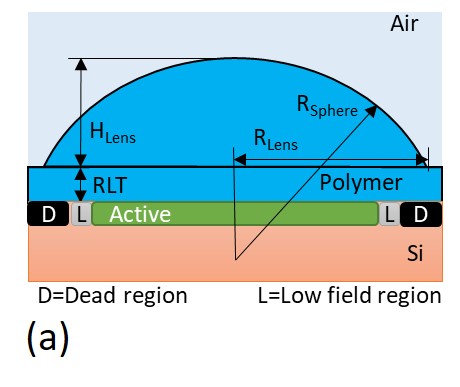}
    \includegraphics[width=0.235 \textwidth, angle=0]{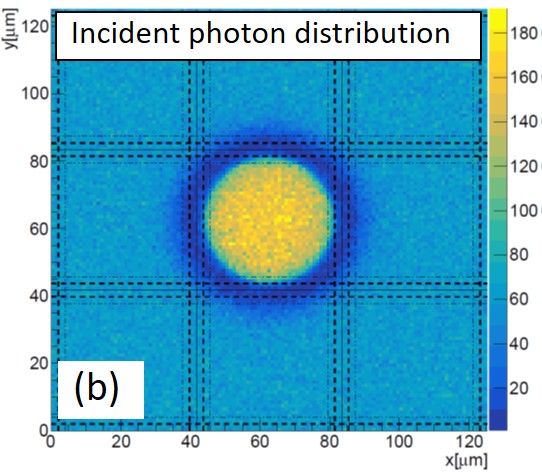}
	\includegraphics[width=0.235 \textwidth, angle=0]{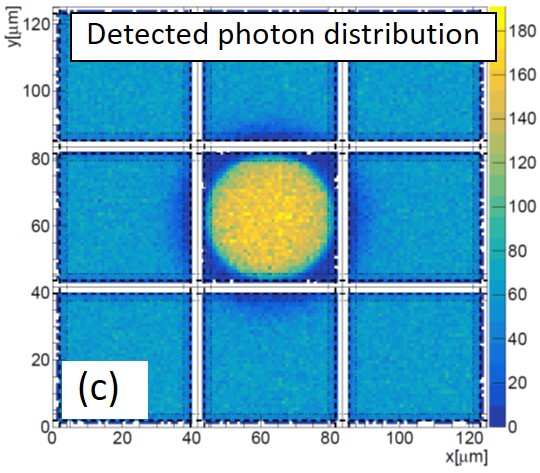}	
	\includegraphics[width=0.235 \textwidth, angle=0]{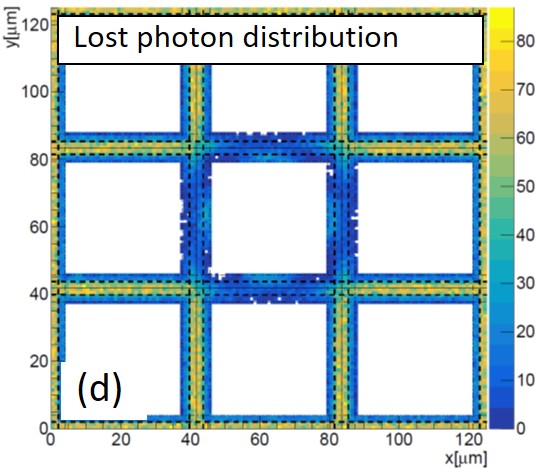}	
	\caption{(a) Illustration of the parameters of the lens and the detection regions of the SiPM pixel. (b) Distribution of the photons arriving at the silicon surface. Note that only a single lens in the center is present in the simulation. (c) Detected photon distribution. (d) Lost photons falling in the dead or partially efficient region. The dead regions with the microlens have a significant lower number of lost photons compared to the dead regions without microlens. The light source angular distribution used for (b), (c) and (d) corresponds to a $230~cm$ long SciFi (polished).} 
	\label{fig:uLens_sim}%
\end{figure}

To optimise the geometry and assess the performance of the microlens implementation, a custom simulation based on a ray tracing Monte Carlo using the ROOT framework~\cite{ROOT} has been developed. The calculation of the effectiveness of the microlens is evaluated for a system of nine pixels with one microlens placed on top of the central pixel. For the actual implementation, not only one in nine but every second pixel is covered with a lens, and therefore the improvement expected from the implementation is scaled with a factor of 4.5 with respect to the simulation. The factor 4.5 corresponds to the mean number of lenses that can be implemented on a grid of $3\times3$ pixels. The simulation results were confirmed in an early phase with a commercial ray tracing simulation software~\cite{zemax} by our industrial partner CSEM. The functionality of the pixel is modelled as active area (centre), dead border regions (outermost borders), and partially efficient low field border region (between dead and active). The GFF is set to a constant value in the simulation but experimental data show lower avalanche trigger probability at the edges~\cite{LF_efficiency_FBK}. 
Therefore, the GFF of the pixel was used to define the dead region size and the low field region was modelled according to the best estimation by the designer at FBK as a band with reduced PDE. 
The improvement is defined as an increase in the number of detected photons. It is equal to the ratio between detected photons with a MLA and a flat coated SiPM device. An illustration of the total incident photons, the detected photons, and the lost photons is given in Fig.~\ref{fig:uLens_sim}. 

The ray tracing simulation allows to calculate the photon path from the light source to the interception point with the silicon. 

\subsection{Light source exit angle distribution}\label{light_source}
\begin{figure}
	\centering 
	\includegraphics[width=0.35 \textwidth, angle=0]{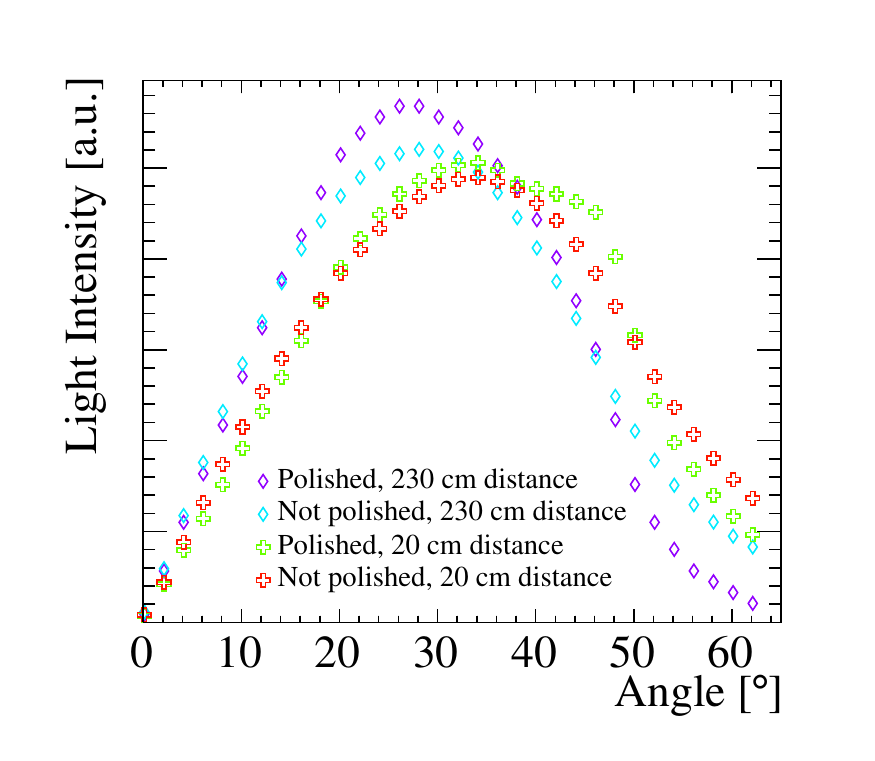}	
	\caption{Fibre exit angle intensity distribution for a double cladded fibre with a numerical aperture NA=0.72 for two distances, defined as the length between the injection point on the fibre and the fibre end. The measurements show the dependency of the exit surface quality (polished or not) and the distance of the injection point from the detector.} 
	\label{fig:exit_angle_dist}%
\end{figure}

For the simulation, two types of light sources were used. One modelling the SciFi exit angle distribution and one with normal incident light as used in our detector characterisation setup. 
To obtain the exit angle distribution from the SciFi, a dedicated setup was used. A 2.5 m long fibre is excited with a UV-LED at a distance of 20 and 230cm from the fibre end. With a camera fixed on a rotary motorised stage, the intensity as a function of angle was measured and normalised. A $3~mm$ aperture placed at a distance of $5~cm$ from the end of the fibre ensures a small angular acceptance. Note that the distribution depends on the injection distance and the quality of the exit surface. In Fig.~\ref{fig:exit_angle_dist} the distribution for a SciFi is given. The long transmission path (long for a plastic optical fibre) reveals a higher attenuation for large angles.  
In the context of the improvement of the SciFi tracking application, the optimisation of the microlens parameters was performed with the SciFi exit angle distribution. SiPMs used in other applications, such as for Cherenkov detectors~\cite{RICH}, astroparticle detectors~\cite{Astro}, and Lidar~\cite{Lidar}, require microlens parameters optimised for specific light sources. In general, these aforementioned applications have narrow normal incident light sources.

\subsection{The optical parameters}
The simulation allowed us to obtain the optimal geometrical parameters for the microlens. The range of parameters was guided (limited) by the limitations of the microlens manufacturing process. The maximal value of the diameter of the lens is the full diagonal of a pixel, but in the mold fabrication process, a minimal distance of $2\,\mu m$ between the lenses is required. The simulation showed that the highest light detection is achieved with a value of 95\% of the pixel diagonal (Table~\ref{tab:result_41_fibre_100}). As this value is expressed as a percentage of the pixel diagonal, it is valid for different pixel sizes. 

The 95\% results in a lens spacing of $2.95\,\mu m$ for a pixel size of $42\,\mu m $ and $2.4\,\mu m$ for a pixel size of $31\,\mu m $. The lens surface curvature is restricted to a spherical shape due to the manufacturing process of the master mould that uses the surface tension during a melting cycle. The height of the lens is restricted via the maximum acceptable inclination angle of the surface. The demolding process can only be reliably performed for angles below $75^{\circ}$ and therefore this limits the maximal lens height. For the simulation, this angle is used for the parameterisation of the lens height. 
The third restriction applies to residual layer height ($RLT$), the thickness of the coating without lenses, as illustrated in Fig.~\ref{fig:uLens_sim}. With a large mould size covering an entire silicon wafer, the minimal $RTL$ that can be reached with this replication process is on the order of 10 to $15\,\mu m $. 
\subsection{Simulation results}
In the simulation, $R_{Lens}$, $H_{Lens}$, and $RLT$ were varied to achieve the highest enhancement in light yield ($LY_{enh}$). The results are summarised in Table~\ref{tab:result_41_fibre_100}. For this study, the dead region of the pixel was uniformly distributed along the edge of the pixel. In addition, an intermediate region $2\,\mu m $ wide at the edge of the active region is implemented modeling the low field region (LFR) which is more pronounced at low $\Delta V$. Here, the efficiency decreases from $100\%$ in Table~\ref{tab:result_41_fibre_100} to $60\%$ in Table~\ref{tab:result_41_fibre_60} where only $RLT$ is varied.  
\begin{table}
\centering 
\caption{Highest light yield enhancement ($LY_{enh}$) achieved with a lens radius of $95\%$ of the pixel diagonal, a maximal lens height, an $RLT=12.5\,\mu m $, and a low-field region with a $100\%$ efficiency. }\begin{tabular}{|c|c|c|} \hline  
 \multicolumn{3}{|c|}{FBK, pixel size $41.7\mu m$, GFF=$81.5\%$, light source SciFi}\\ \hline  
 $H_{Lens} [^\circ]$ & $RLT [\mu m]$ & $LY_{enh} [\%]$\\ \hline 
 \multirow{4}{*}{65} & 10  & 13.35\\  
 & 12.5 & 15.15\\  
 & 15   & 15.52\\  
 & 17.5 & 15.13\\ \hline 
 \multirow{4}{*}{70} & 10   &  15.88\\  
 & 12.5 & 17.22\\  
 & 15   & 16.83\\  
 & 17.5 & 16.29\\ \hline 
 \multirow{4}{*}{75} & 10   & 18.17\\  
 & \bf{12.5} & \bf{18.59}\\  
 & 15   & 17.12\\  
 & 17.5 & 16.12\\ \hline 
\end{tabular}

\label{tab:result_41_fibre_100}
\end{table}

\begin{table}
\centering 
\caption{Low-field region $2\,\mu m$ wide at a $60\%$ efficiency models the reduced $P_{AT}$ for the operation at low $\Delta V$ for a lens with a radius of $95\%$ of the pixel diagonal and a lens height of $75 ^\circ$. }
\begin{tabular}{|c|c|} \hline  
 \multicolumn{2}{|c|}{FBK, pixel size: $41.7\mu m$ GFF: $81.5\%$, light source SciFi}\\ \hline  
 $RLT [\mu m]$ & $LY_{enh} [\%]$\\ \hline 
 10   & 24.54\\  
 \bf{12.5} &\bf{24.57}\\  
 15   & 23.68\\  
 17.5 & 22.11\\ \hline 
\end{tabular}
\label{tab:result_41_fibre_60}
\end{table}

\subsection{Simulation beyond the current detector implementation and for normal incident light}
In addition to the $41.7\,\mu m$ pixel from FBK with a very high $GFF=81.5\%$ and the large exit angle distribution light source (SciFi), other cases were studied. Particularly interesting is the optimisation of a narrow and normal incident light source. A list of parameters and simulated $LY_{enh}$ is given in Table~\ref{tab:result_param_others}. There, a comparison of different pixels shown in Fig.~\ref{fig:pixel_layout} is given. 
The larger (SciFi) exit angle distribution leads to optimal parameters with small $RLT$ and high lenses ($H_{Lens}=75^\circ$). With a narrow and normal incident light source, thicker $RLT$ and flatter lenses can be efficient. Smaller pixels and in general pixels with lower $GFF$ have the potential for a larger improvement. To simplify the comparison, we assume a $2\,\mu m$ low field region with $R_{Lens}=95\%$ in all cases. 
\begin{table}
\centering
\caption{Detectors with different pixel sizes from HPK and FBK, all with a $RLT=12.5\,\mu m$ and $H_{Lens}=75^\circ$, are studied with a narrow normal incident light and a SciFi angular distribution.}\begin{tabular}{|c|c|c|} \hline  
 $GFF [\%]$ & Light source & $LY_{enh} [\%]$\\ \hline 
 \multicolumn{3}{|c|}{FBK, pixel size: $41.7\mu m$  }\\ \hline  
 \multirow{2}{*}{81.5} & SciFi  &  24.57 \\  
  &  Narrow & 27.91 \\  \hline 
 \multicolumn{3}{|c|}{HPK, pixel size: $41.7\mu m$  }\\ \hline  
 \multirow{2}{*}{68.5} & SciFi & 38.98 \\  
  & Narrow & 45.93 \\  \hline 
 \multicolumn{3}{|c|}{FBK, pixel size: $31.3\mu m$  }\\ \hline  
 \multirow{2}{*}{77.7} & SciFi  &  27.51 \\  
  &  Narrow & 39.87 \\  \hline 
\end{tabular}

\label{tab:result_param_others}
\end{table}

\subsection{Optical coupling}
The microlens principle is based on the refraction of light between the air and the lens material. Any application that requires optical glue or grease between the light source and the photodetector (e.g. LYSO crystals glued to SiPM surface) will not profit from the microlens, whereas systems where a narrow exit angle distribution of the light source is present have high potential for enhancing the PDE with MLAs.

\section{The implementation of prototypes}
In view of the development of a new multichannel SiPM array for LHCb Upgrade 2, a dedicated implementation was defined to evaluate microlens enhancement. In addition to the constraint of the geometrical aspects of the active surface and the chip dimensions, additional specific alignment markers were added. To reach thin $RLT$, the original 8-inch wafers were diced into pieces of $2\times2$ reticles of $32\,mm\times 32\,mm$ in size. The detectors with $41.7\,\mu m$ pixel size were chosen for this study and two types of samples were prepared, flat layer (uniform epoxy layer) and MLA. 

\subsection{Silicon}
\begin{figure}
	\centering 
	\includegraphics[width=0.24 \textwidth, angle=0]{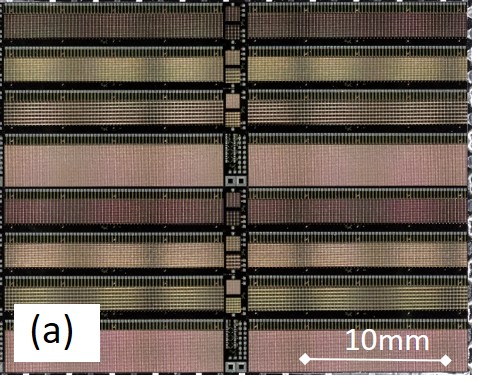}	
	\includegraphics[width=0.230\textwidth, angle=0]{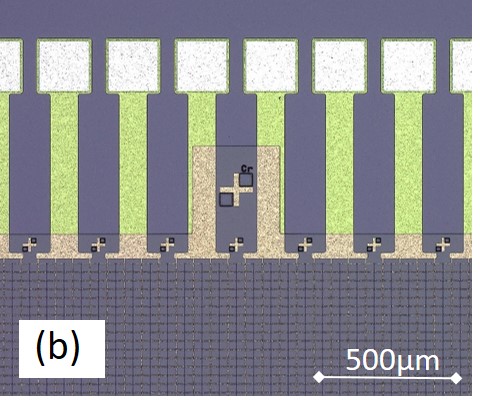}	
	\caption{(a) shows four reticles with each four dies with different pixel sizes. The MLA replication is applied to this size of such wafer pieces. (b) shows the bond pads, alignment markers, and a flat layer. The flat layer covers the active area but not the bond pads (slight orange color).} 
	\label{fig:flat_layer_coated}%
\end{figure}

\begin{figure}
	\centering 
	\includegraphics[width=0.24\textwidth, angle=0]{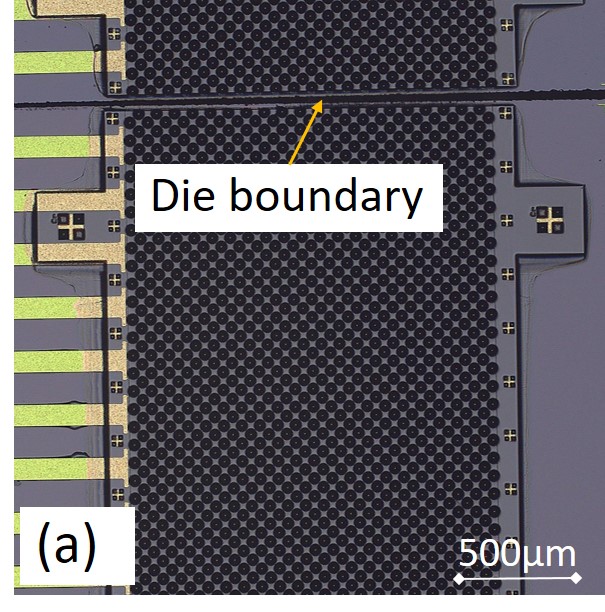}	
	\includegraphics[width=0.30\textwidth, angle=0]{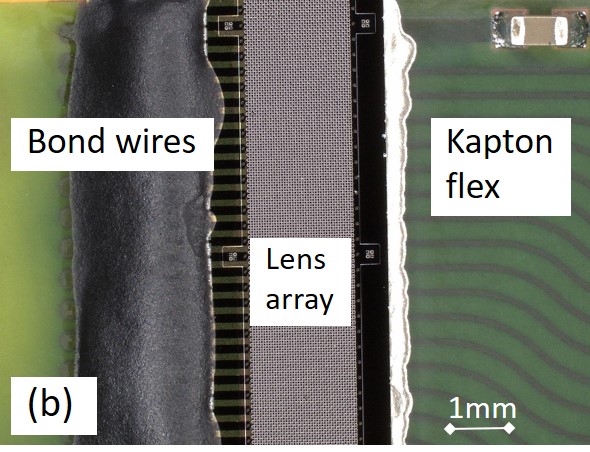}	
	\caption{(a) two microlens array dies bonded on one module. (b) the arrays packaged on the Kapton flex circuit with the protected bond wires and the traces on the readout circuit.} 
	\label{fig:flex_assembly}%
\end{figure}

A custom implementation of a SiPM array was produced by FBK based on the NUV-HD technology~\cite{FBK_NUV}. This technology features a high $GFF$ of $81.5\%$ for a pixel size of $41.7\,\mu m$ and can be operated in a wide $\Delta V$ range. For this production, three different pixel sizes were included in a reticle, but for the MLA implementation, only the $41.7\mu m$ chip type was used. The silicon dies are multichannel arrays with 64 channels with a size of $1.62\,mm \times 250\,\mu m$ per channel and a total size of $16\,mm \times 1.62\,mm$. For an MLA replication, 4 dies were produced in a single replication process, but since other dies were also present in the reticle, the total replication area on a $2\times2$ reticle is $32\,mm \times 32\,mm$. The connectivity is implemented with a common cathode on the back and bond pads on the top at a distance of $700\,\mu m$ from the active surface to allow for a bond wire protection with glue. The MLA layer deposition is restricted to the active surface but leaves the surface of the bond pad uncoated (Fig.~\ref{fig:flat_layer_coated}). 
The deposit of the MLAs and flat layers was performed on eight dies each. The optical properties of the samples were evaluated before dicing and packaging. The package is a bare die bonding of two dies on a flexible Kapton circuit with a ceramic stiffener and temperature sensor on the backside. The assembled module can be seen in Fig.~\ref{fig:flex_assembly}. 
For electrical tests, these samples were characterised with a tunable monochromatic light source for the test with narrow normal incident light as well as with a SciFi mat (array of fibres) that provides the specific angular distribution discussed in Sect.~\ref{light_source}. 

\section{Visual evaluation}
Before the electrical properties were measured, a visual inspection of the characteristics of the different samples was performed. Visual inspection was performed in several steps in mould production and in the replication stage.

\subsection{Evaluation of the optical parameters}
\begin{figure}
	\centering 
	\includegraphics[width=0.3 \textwidth, angle=0]{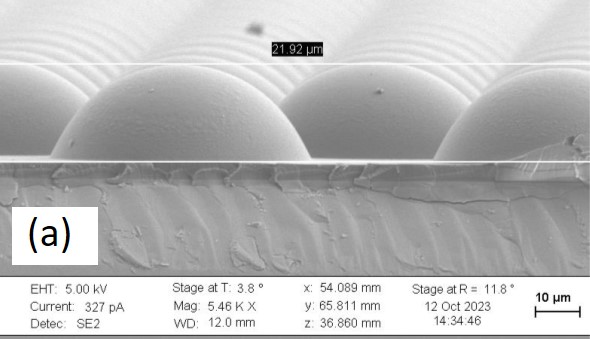}	
	\includegraphics[width=0.3 \textwidth, angle=0]{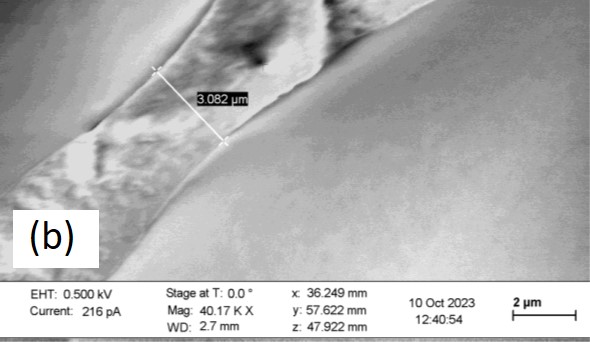}	
	\caption{SEM images showing (a) the cross-section of the test structure allowing to measure the lens height, and (b) the distance between two lenses seen from the top.} 
	\label{fig:sem_image}%
\end{figure}

The MLAs have been evaluated at different production steps. Before the deposition of the lens on the silicon, a test replication on a glass wafer plate and a scanning electron microscope (SEM) image were produced to cross-check the obtained dimensions. In Fig.~\ref{fig:sem_image} a lens height of $21.9\mu m$ is measured, where the nominal design parameter was $21.5\mu m$. For the distance between lenses, the SEM image is produced from the top. Note that the characterisation of transparent non-flat optical components is difficult with an optical microscope as variable focal planes are required.  
The statistical measurements of the structures show an excellent correspondence for $R_{Lens}$ and $H_{Lens}$, as can be seen in Table~\ref{tab:result_param}. $R_{Lens}$ is measured indirectly through the measured minimum distance between two lenses (minimal gap), and the height of the lens is the height over the residual plane. The next steps of the verification of the implementation are the evaluation of the $RLT$ and the alignment of the replicated pattern with respect to the structures on the silicon. Unlike the optical parameter of the lens, these properties are not fixed for a given replication mould but depend on the replication process. It is therefore useful to measure for every chip the $RLT$ and validate the alignment at several points on the surface.  
\begin{table}
    \centering
    \caption{List of nominal and measured values for optical parameters of the master mold and parameters of a specific sample implementation. For the misalignment only the maximal deviation in the diagonal direction is given. }
    \begin{tabular}{|c|c|c|} \hline
      Parameter     & Nominal     & Measured (min-max)\\ \hline
      Minimal gap   & $2.95\,\mu m$ & $2.6-3.5\,\mu m$ \\ \hline
      Lens height   & $21.5\,\mu m$ ($75^\circ$)& $21.7-23.5\,\mu m$\\ \hline
      RLT microlens & $12.5\,\mu m$ & $11.1-13.7\,\mu m$\\ \hline
      RLT flat      & $12.5\,\mu m$ & $8.9-10.8\,\mu m$\\ \hline
      Misalignment  & 0           & $2.0\,\mu m$\\ \hline
    \end{tabular}
    \label{tab:result_param}
\end{table}
The extracted optical parameters match the nominal values very well. As the simulation results in Table~\ref{tab:result_41_fibre_100} show, a small difference for $H_{Lens}$ does not lead to important changes in efficiency. Misalignment is the most important factor that can decrease the efficiency of the MLA. Thus, a dedicated alignment and control system was implemented. 

\subsection{Alignment}
\begin{figure}
	\centering 
	\includegraphics[width=0.39 \textwidth, angle=0]{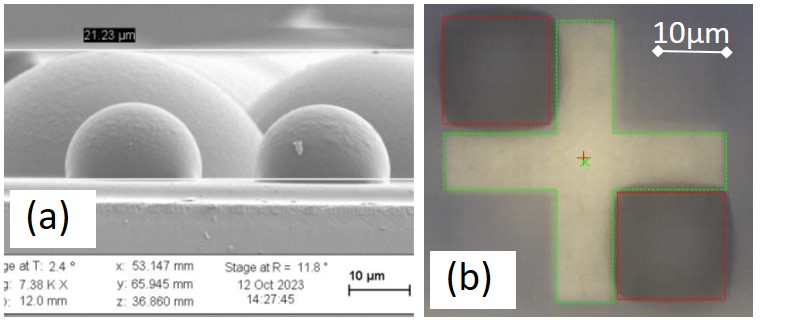}	
	\includegraphics[width=0.36 \textwidth, angle=0]{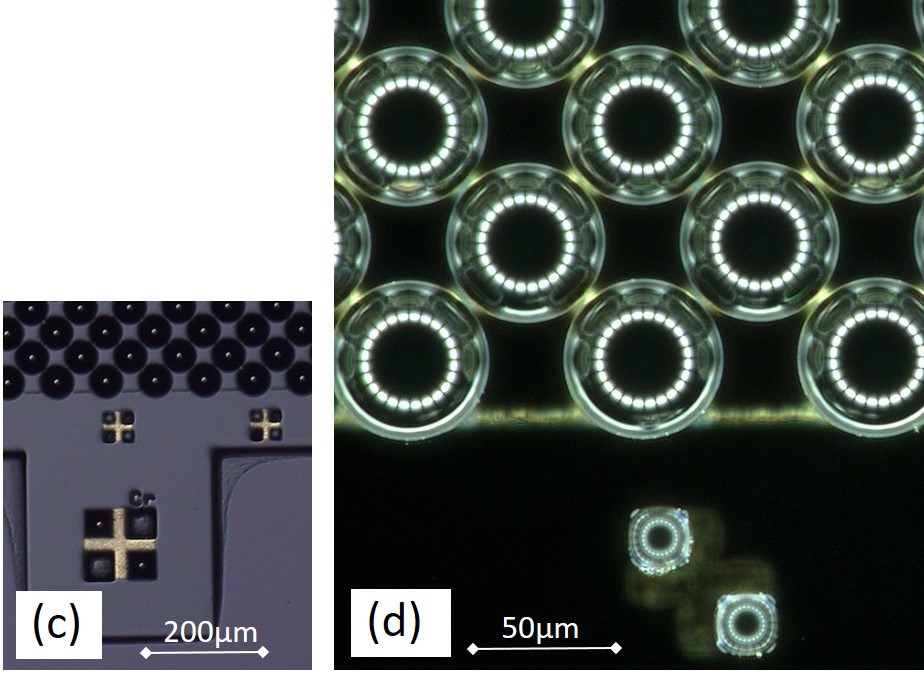}	
	\caption{(a) SEM image of two smaller size microlenses placed on top of the alignment marker. (b) Close view of the alignment markers filled with the microlenses. (c) Coated region with microlenses with coaxial illumination. (d) With a segmented diffuse light source, the reflection on each microlens is visible and illustrates the curved surface. In this mode, the flat regions between lenses appear black and the pixel structure (routing lines) is made invisible due to the lenses.} 
	\label{fig:sem_alignment}%
\end{figure}

The MLA replication is performed with a mold that requires a precise alignment with respect to the structures on the silicon. Markers were added to the silicon not only for the alignment process during replication but also for quality assurance after MLA deposition. Microlenses with reduced diameter and height were integrated into the mold and placed within the cross corners of the alignment marker. The misalignment was measured with an optical microscope in every corner of the MLA through an automated process. The precision is typically better than $1\,\mu m $ in both the horizontal and vertical directions. A maximum misalignment of $2\,\mu m $ in diagonal was observed, but for many samples the misalignment was of the order of $1\,\mu m $ in diagonal.  In Fig.~\ref{fig:sem_alignment} the relative position of the microlens implementation with respect to the alignment markers is illustrated.

\section{Comparison with measurements}
The evaluation of the MLA is based on a relative comparison between MLA enhanced and flat layer samples. The expected differences are a change in PDE and external optical cross-talk\footnote{Generated by photons from a primary avalanche exiting the silicon, reflected on the entrance window and producing a sub-sequent avalanche.} due to the different surface structures, although also the SPTR is expected to change. To extend the variation of samples, we also measured bare silicon from the identical fabrication process. The detectors are 128-channel arrays which yield data from many channels. This reduces the systematic uncertainties caused by manufacturing differences and small defects in the microlens surface. However, the alignment is linked to a 64-channel die and is expected to affect all channels on the same die. 

\subsection{Breakdown voltage and gain measurement}
\begin{figure}
	\centering 
	\includegraphics[width=0.235\textwidth, angle=0]{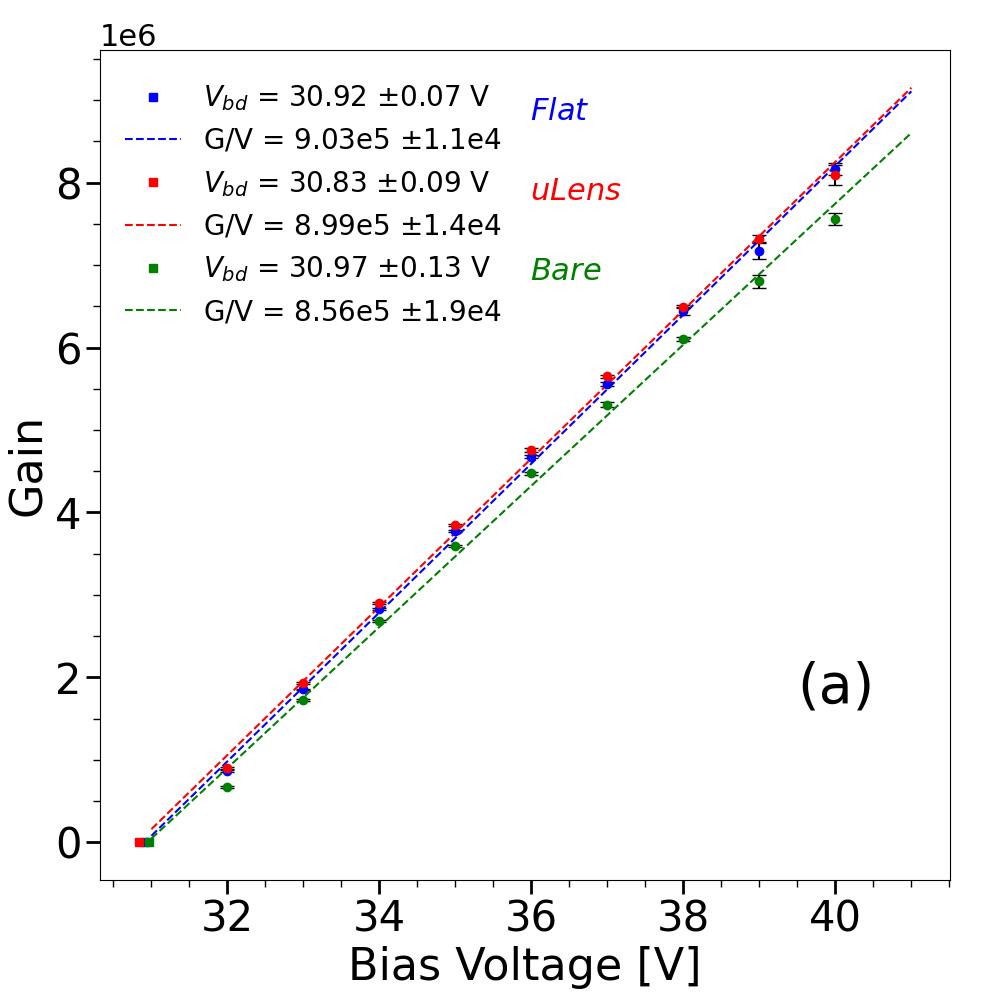}	
	\includegraphics[width=0.235\textwidth, angle=0]{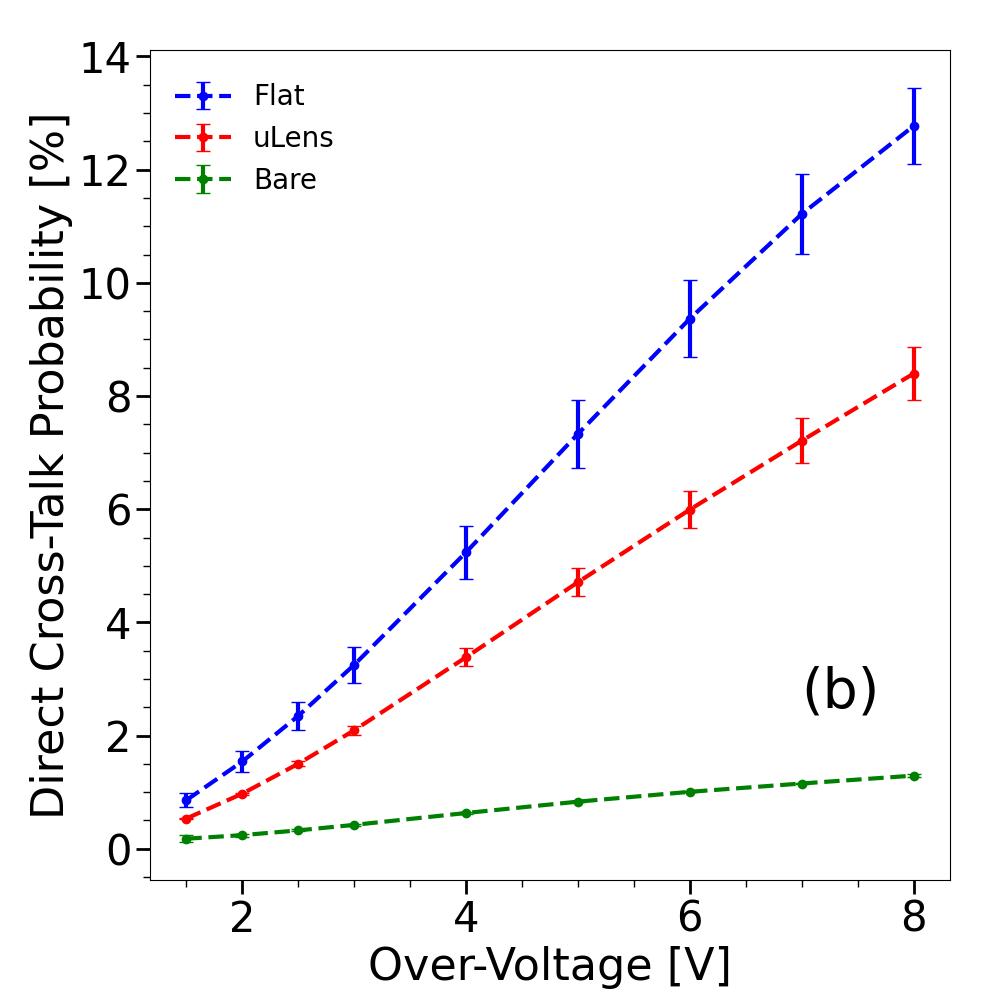}	
	\caption{(a) $V_{bd}$ is calculated by the extrapolation of the linear fit to the intercept with zero gain allowing all measurements to be presented at equal $\Delta V$. The gain is calculated from the slope of the linear fit and is given for bare, flat and microlens coated samples. Note that due to the coating, the gain is slightly increased. (b) a large increase of the cross-talk is seen for the coated samples. The bare sample is expected to have no external cross-talk but only internal. The error bars show the error of four measured identical channels.} 
	\label{fig:vbd_cross_talk}%
\end{figure}

The breakdown voltage is measured for every channel and, although they are integrated on the same silicon die, a spread is seen due to manufacturing differences. Over 128 channels, it is typically in the range of $100-300\,mV$. The temperature dependence of $V_{BD}$ for this technology was measured to be $32\,mV/K$ at room temperature. An integrated Pt1000 temperature sensor on the detector module gives access to precise temperature monitoring. With the temperature in the lab stabilised by air conditioning, the variations were smaller than $\pm 1\,K$. To determine $V_{BD}$ and the gain (G), the bias voltage was increased in steps of $1\,V$ and the current $I_{Bias}$ recorded. At the same time, pulse rates at thresholds of 0.5 and 1.5 photo-electron amplitude (PE) ($R_{0.5}$, $R_{1.5}$) were recorded with a high-speed oscilloscope. The gain at every $V_{BD}$ point is calculated as the ratio between the current and the sum of $R_{0.5\,PE}$ and $R_{1.5\,PE}$ and is divided by the elementary charge (e) such that $G=I_{Bias}/(R_{0.5PE}+R_{1.5PE})/e$. A linear fit is used to extrapolate $V_{BD}$ and the slope is extracted from the fit to determine the gain. The results are shown in Fig.~\ref{fig:vbd_cross_talk}. The measurement is performed with light illuminating the detector to enhance the precision of the current measurement but with a sufficiently low pulse rate such that the random overlap of pulses in time is negligible. The measurement is repeated for every wavelength, and the error is given as the error bars in the plot. 

\subsection{Cross-talk and after-pulse}
The ratio of the rates at thresholds $1.5PE$ and $0.5PE$ defines the cross-talk probability $P_{x-talk}=R_{1.5PE}/R_{0.5PE}$). The data were recorded from four detector channels each, and the error bars indicate the spread between channels. $P_{x-talk}$ is increased by the flat and microlens coating. With an average thickness of $10\mu m$ for flat surfaces and $12.4\mu m$ for microlens coating, and a pixel size of $41.7\mu m$, the external optical cross-talk is expected to increase over bare silicon. Even such a thin ($10-20\mu m$) coating increases the probability of external optical cross-talk. A reduction from $13\%@8V$ for flat to $8.5\%@8V$ for microlens is observed, with almost identical $RLT$. This corresponds to a reduction in external optical cross-talk of $40\%@8V$. The cross-talk as a function of $\Delta V$ is given in Fig.~\ref{fig:vbd_cross_talk}. At room temperature and $\Delta V<10\,V$, the after-pulse probability is below $0.5\%$. As the main results for PDE and light yield improvement are calculated as ratios, it is well justified to neglect after-pulse corrections.

\subsection{PDE with narrow, normally incident light source}
The three different detectors were characterised with a narrow, normally incident light source. A Xenon lamp and a tunable monochromator are used to produce a monochromatic light beam that is coupled into a multimode glass fibre and homogenised with a microlens based homogeniser\footnote{https://www.lightsource.tech/en/accessories/homogenizer}. The detector (SiPM array) together with a reference photodiode (PD) is mounted at a distance of $350\,mm$ on a motorised linear stage, which allows for centering of the detectors with respect to the light beam. A homogeneity better than $\pm 0.5\%$ is achieved over the reference PD of $100\,mm^2$ whereas for the small SiPM channel $0.408\,mm^2$ a much better value is reached. With the calibration curve for the PD (sensitivity as a function of wavelength) and the ratio of the surface between the two detectors, the absolute PDE is calculated with the recorded pulse rate $R_{0.5\,PE}$ corrected by the cross-talk rate $R_{1.5\,PE}$ and the dark pulse rate. The dark rates are typically $20-100$ times smaller than the rates with light.
For the evaluation of the MLA, we provide absolute and relative changes in PDE, where the latter has a lower systematic uncertainty, as it is not dependent on the absolute PD calibration nor the homogeneity of the illumination. 

\begin{figure}
	\centering 
	\includegraphics[width=0.235 \textwidth, angle=0]{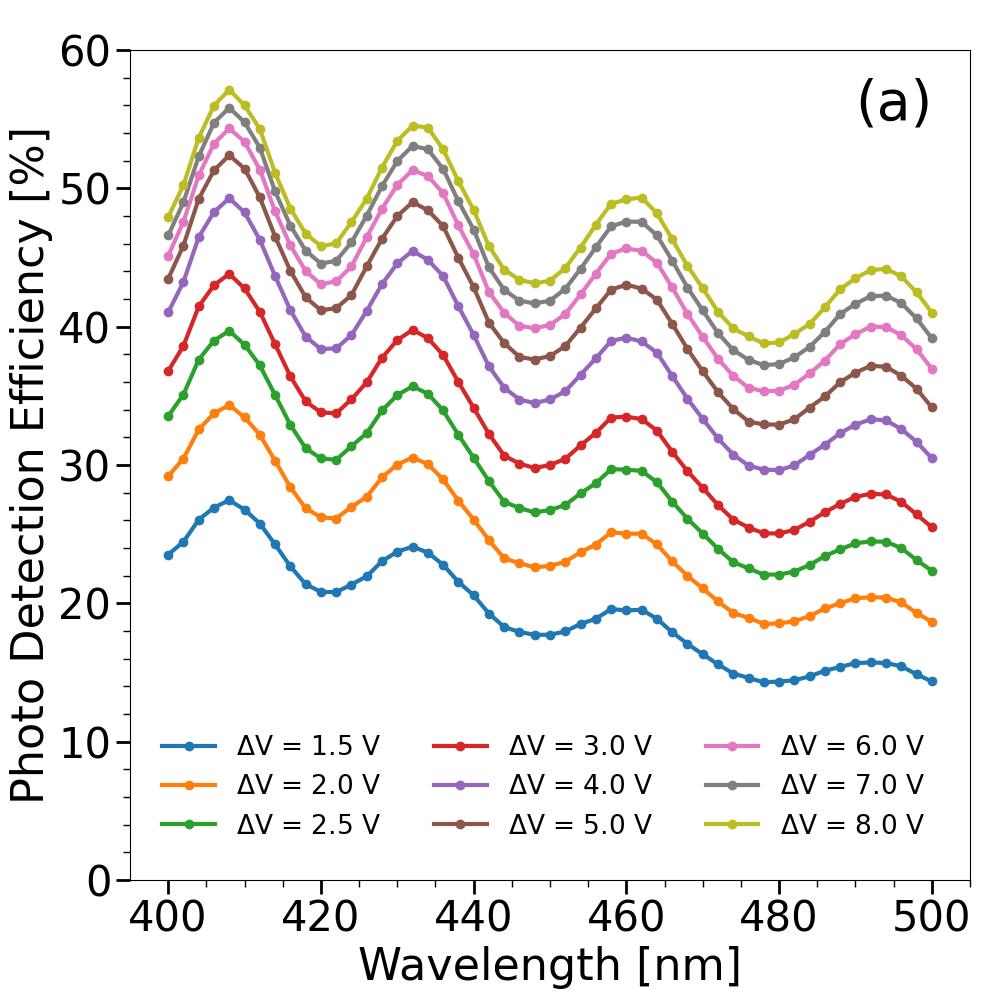}	
	\includegraphics[width=0.235 \textwidth, angle=0]{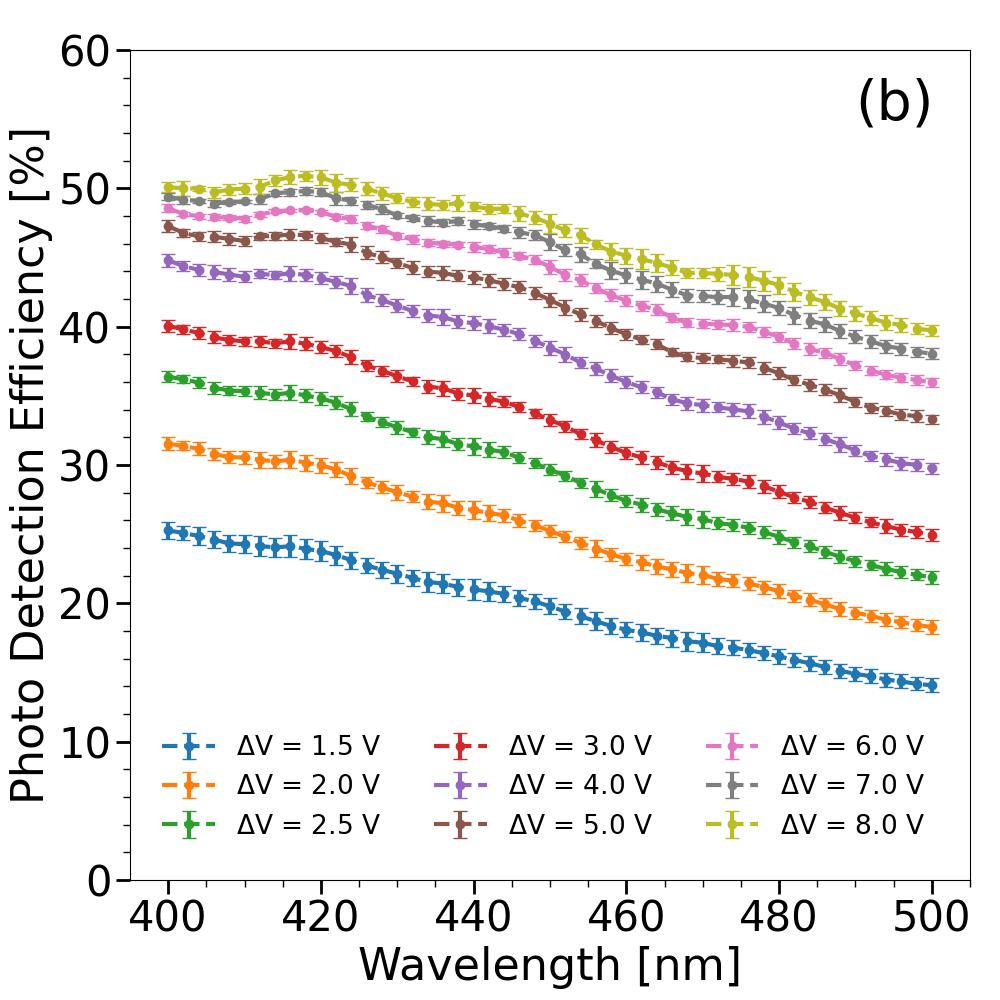}	
	\includegraphics[width=0.235 \textwidth, angle=0]{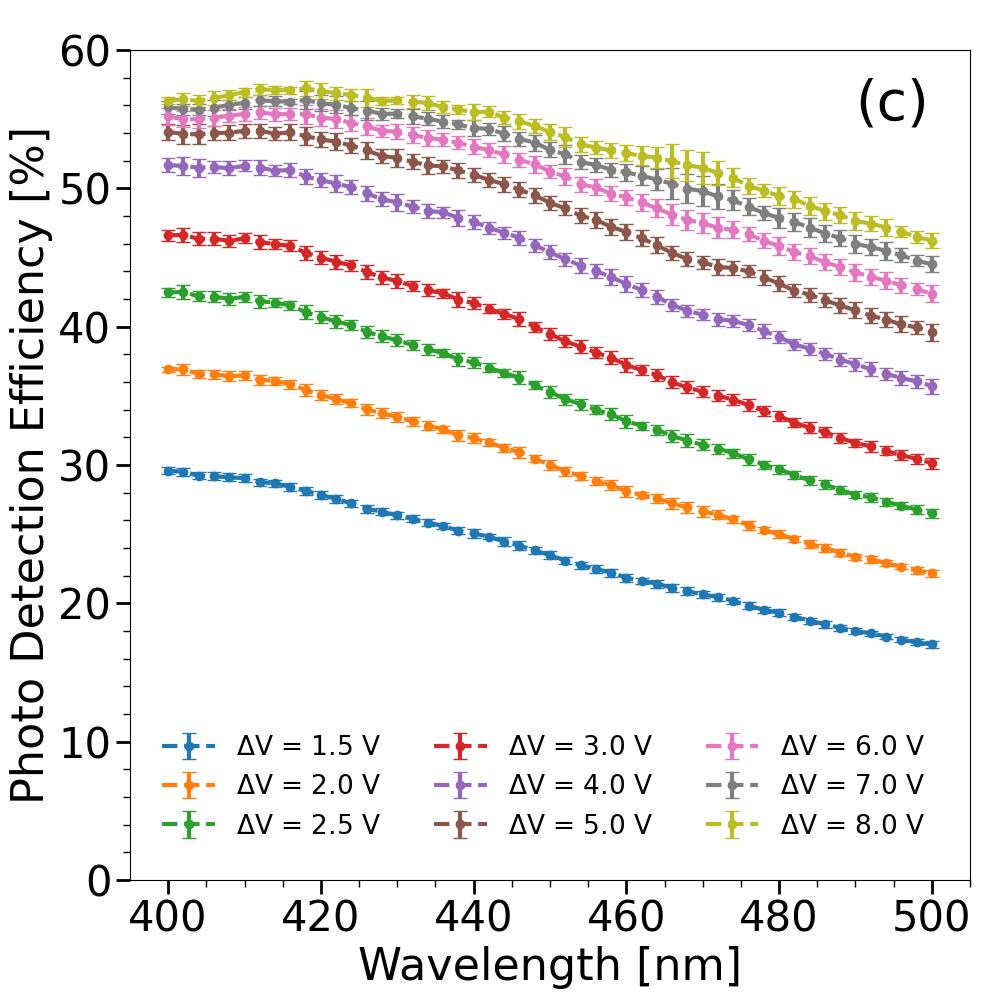}	
	\includegraphics[width=0.235 \textwidth, angle=0]{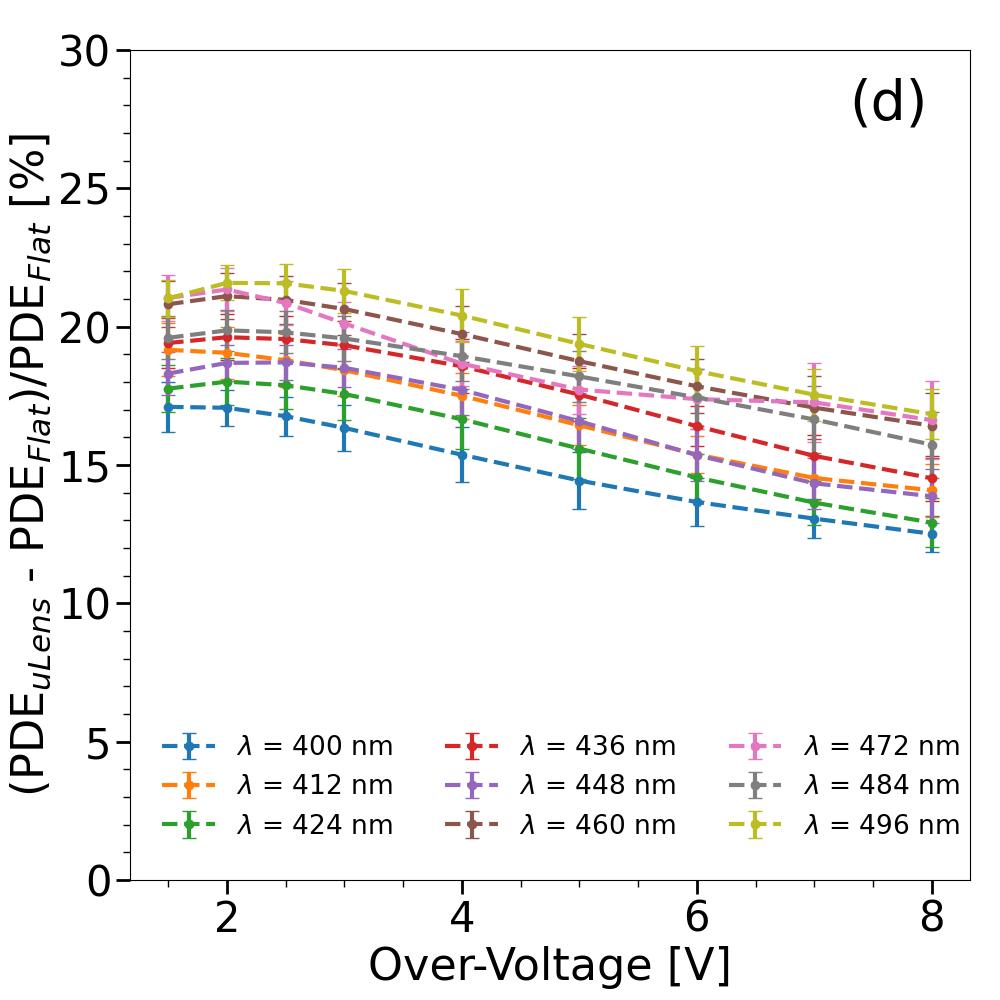}	
	\caption{Results of PDE measurements with narrow normal incident light. (a) PDE for bare surface, interference pattern generated by the anti-reflecting coating are well visible. (b) PDE of flat polymer coated detector. (c) PDE of microlens-enhanced detectors. (d) PDE increase for microlens over flat layer coating. A maximum of 22\% is reached at low $\Delta V$.} 
	\label{fig:pde}%
\end{figure}
Measurements are performed in a fully automated setup where limited temperature control (by air conditioning) is used. The $V_{bd}$ is calculated for each measurement for every channel, and the PDE is given as a function of a $\Delta V$ in~Fig.\ref{fig:pde}. The results show only small fluctuations between channels on the same silicon dies. The error bars indicated in the plots are RMS for four channels. The wavelength dependence can be explained by a change in the refractive index of the microlens polymer material, which is between $1.53-1.51$ in the wavelength range between $400-500\,nm$. The increase in PDE is lower for higher $\Delta V$, which is in accordance with the effect expected due to the low field region of the pixel. The simulated increase with normal incident light for this implementation is $27.91\%$ at low $\Delta V$ (with a low field region) and $21.11\%$ at high $\Delta V$ (without a low field region). This can be compared with the measurement result of $22\%$ at low $\Delta V$ and $17\%$ at high $\Delta V$.  

\subsection{Comparison with SciFi light source}
\begin{figure}
	\centering 
	\includegraphics[width=0.35\textwidth, angle=0]{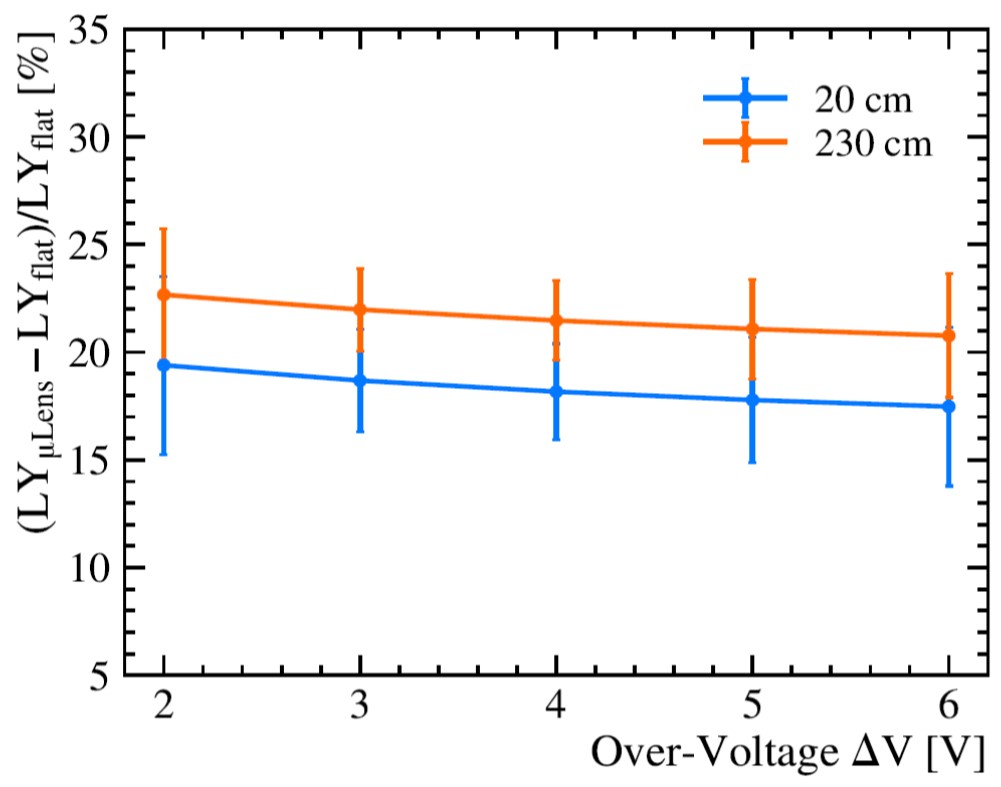}	
	\caption{Measurement of a SciFi fibre mat (light source) with a total length of $245\,cm$. The microlens-enhanced detectors compared to the flat layer devices show up to 23\% LY increase. The error bars shown are due to the fit uncertainty of the signal amplitude distributions. The measurements shown are for two different injection points with a distance of $20\,cm$ and $230\,cm$ from the photodetector.} 
	\label{fig:LY}%
\end{figure}

The microlens parameters were optimised for the SciFi as a light source and represent one of the goals of the MLA project. The test with the SciFi was performed at CERN SPS with pions at a momentum of $p=180\,GeV/c$. The SciFi module was arranged to perform a longitudinal scan. The distance between the particle beam and the photodetector was varied between $20\,cm - 230\,cm$ and the light intensity of the normal incident particles was measured. With a Gaussian fit, the median is extracted from the characteristic Landau distribution introduced by the energy deposit fluctuations. This value is called the light yield (LY) of the SciFi tracker. As a consequence of the mirror placed at the far end of the SciFi mat at a distance of $245\,cm$ from the SiPM, the detected light is composed of a direct component and a mirrored component. At a distance of $20\,cm$, the mirrored component is reduced to about $20\%$ due to attenuation. A fibre mat of $130\,mm$ width was equipped with four flat layer dies and four microlens dies, 256 channels each. In Fig.~\ref{fig:LY} the LY is plotted as a function of $\Delta V$ and given for two distances from the detector. The LY measurement has been performed after a gain calibration with a light injection system for each acquisition channel.   

For the SciFi light distribution for a propagation length of $230~cm$, the simulated LY enhancement is $24.57\%$ for low $\Delta V$ and $18.59\%$ for high $\Delta V$, and the measured values with SciFi are $22.5\%$ for low $\Delta V$ and $20.5\%$ for high $\Delta V$. Note that the simulation was performed with a light exit angle distribution corresponding to a $230~cm$ long fibre.

\section{Conclusion} 

In this work, we present a novel implementation of MLA with the objective of improving the PDE of SiPMs. The implementation was guided by the technology limitation of the MLA deposit by a mold and a UV-curable polymer. To optimise the most important parameters, the geometry of the SiPM pixel was simulated with three regions: the dead region, the low field region representing a reduced PDE region at low $\Delta V$ and the fully active region. The MLA consists of spherical lenses placed in a checkerboard pattern aligned with the pixel structure. The simulation enables the optimisation of the parameters for lens diameter, lens height, and residual layer thickness. MLAs at the wafer level were deposited on SiPM multichannel arrays with $41.7\,\mu m$ pixel size from FBK and compared with bare and flat layer coated devices. The performance was measured in the lab by injecting monochromatic narrow normal incident light and in a fibre tracker module with high energetic particles. The results are presented in detail as a function of wavelength and $\Delta V$ for the laboratory measurement and as a function of injection distance and $\Delta V$ for the SciFi measurement.\par

In summary, at $\Delta V=2\,V$, an increase of $22\%$ is observed for the narrow normal light source and  $17\%$ at $\Delta V=8V$. For the SciFi light source, the increase is $22.5\%$ and $20.5\%$ at $\Delta V=2\,V$ and $\Delta V=6\,V$, respectively. These values correspond to an overestimation of $10-15\%$ by simulation. As the implementation depends on the precise alignment of the MLA and the silicon structure as well as on the optical parameters of the implementation, the results of simulation and measurements can be considered in good agreement.
\par
In addition to the higher PDE, MLA-enhanced detectors also show a reduction of $40\%$ of external cross-talk compared to a flat coated detector. An improvement in timing performance in SPTR is expected due to the partial masking of the border region (deflection of the photons towards the pixel centre by the lens), which was reported in other works and seen in preliminary measurements in our lab.
\par
The optical transparency of the less than $50\mu m$ thick MLA layer, in a wavelength range between $350nm$ and $700nm$, was measured before and after irradiation. No measurable change was observed with a neutron equivalent fluence of the expected radiation level in the LHCb upgrade 2 ($3.0 \times 10^{12}~{\rm 1\,MeV\,n_{eq}/cm^2}$) and an ionising dose of $1kGy$. Further improvements by adding an anti-reflecting coating (ARC) to the polymer surfaces of the MLA can potentially decrease the Fresnel reflection loss. A prototype with optimised coating for the emission spectral range of the SciFi light source is foreseen. For low $\Delta V=2V$, the prospects of MLA enhanced SiPMs with SciFi tracker technology are excellent. With the latest implementation of HPK $41.7\mu m$ pixel size, a simulated improvement of $39\%$ is expected. The smaller $31.3\mu m$ pixel size from FBK $28\%$ is predicted to improve.  The smallest pixel size for the replication method used is currently $16~\mu m$ and is limited by the precision of alignment and the thickness of the residual layer. 
These two configurations are currently in preparation for MLA deposition, and new results will be available by the end of 2024.\\ 

\bibliographystyle{ieee}

\end{document}